\documentstyle[12pt]{article}

\newcommand{\wt}{\widetilde}
\newcommand{\het}{{(het)}}
\newcommand{\iia}{{(IIAa)}}
\newcommand{\iib}{{(IIAb)}}
\newcommand{\SG}{{(SG)}}
\newcommand{\be}{\begin{equation}}
\newcommand{\ee}{\end{equation}}
\newcommand{\ben}{\begin{eqnarray}\displaystyle}
\newcommand{\een}{\end{eqnarray}}
\newcommand{\refb}[1]{(\ref{#1})}

\newcommand{\sectiono}[1]{\section{#1}\setcounter{equation}{0}}

\font\mybb=msbm10 at 11pt
\def\bb#1{\hbox{\mybb#1}}
\def\ZZ {\bb{Z}}

\def\CC {\bb{C}}

\font\mybc=msbm10 at 10pt
\def\bc#1{\hbox{\mybc#1}}
\def\ZZZ {\bc{Z}}

\begin{document}
\thispagestyle{empty}
{\hfill CALT-68-2007

\hfill TIFR/TH/95-31

\hfill hep-th/9507027

\vspace {1.0cm}}
\parskip=9pt     
\begin{center}{\Large \bf Type IIA Dual of the\break Six-Dimensional
CHL Compactification}
\end{center}
\bigskip
\bigskip
\centerline{\large John H. Schwarz}
\centerline{\large California Institute of Technology}
\centerline{\large Pasadena, CA 91125}
\medskip
\centerline{and}
\medskip
\centerline{\large Ashoke Sen }
\centerline{\large Tata Institute of Fundamental Research }
\centerline{\large Homi Bhabha Road, Bombay 400005, INDIA }

\bigskip
\bigskip

\begin{abstract}

We propose a candidate for the
dual (in the weak/strong coupling sense) of the six-dimensional
heterotic string compactification constructed recently by
Chaudhuri, Hockney and Lykken. It is a type IIA string theory
compactified on an orbifold $K3/\ZZZ_2$, where the $\ZZZ_2$ action
involves an involution of $K3$ with fixed points,
and also has
an embedding in the U(1) gauge group associated with the
Ramond-Ramond sector of the type IIA string theory. This
introduces flux of the U(1) gauge field
concentrated at the orbifold points. This construction provides
an explicit example where the dual of a super-conformal field theory
background of the heterotic string theory is not a standard
super-conformal field
theory background of the type IIA string theory.

\end{abstract}

\vfil\eject


It has been conjectured recently that various heterotic string
compactifications are nonperturbatively
equivalent to type IIA string compactifications. The first example
of this kind is the equivalence between heterotic string theory
compactified on a four-dimensional
torus and type IIA string theory compactified on
the $K3$ surface\cite{HT,VAFA,DUFF,WITTEN,AS,HS,HT2,VW}.
The massless sectors of both of these theories correspond to non-chiral
$N=2$ supergravity in six dimensions with a gauge group of rank
24. Toroidal compactification to four dimensions then gives dual
theories with $N=4$ supersymmetry. More recently, examples of
dual pairs of theories in four dimensions with $N=2$ supersymmetry have been
found\cite{KACHRU,FHSV}. The conjectured equivalences of
these theories require some highly non-trivial identities to be
satisfied, many of which have been
verified explicitly\cite{GAVAN,LOUIS,KLEMM}.

In another interesting development,
Chaudhuri, Hockney and Lykken (CHL) have constructed
new examples of heterotic string compactification with $N=4$
supersymmetry in four dimensions or
$N=2$ supersymmetry in six dimensions\cite{CHL}. The original
construction was formulated
in terms of fermionic world sheet variables, but
recently Chaudhuri and Polchinski have constructed one of
these models as an asymmetric orbifold\cite{CP}. In simple terms, this
model
corresponds to a $\ZZ_2$ orbifold of the toroidally compactified heterotic
string theory, where the $\ZZ_2$ acts by exchanging
the two $E_8$ components of the momentum lattice, together with a shift by
half a period along one
of the compact directions. The effect of this $\ZZ_2$ modding is to remove
eight of the $U(1)$ gauge bosons from the spectrum, thereby reducing the
rank of the gauge group by eight. This construction does not
require that the $E_8\times E_8$ gauge group is unbroken before the
$\ZZ_2$ modding, but only that the two
$E_8$'s are broken in an identical
manner so that the resulting lattice still has a $\ZZ_2$ automorphism.
This allows us to work at a point in the moduli space where the unbroken
gauge group is just a product of the $U(1)$ factors.

The question that we wish to address is whether there are type IIA string
compactifications that are non-perturbatively
equivalent to the CHL compactification of the heterotic string.
Since the CHL construction can be understood as an orbifold of the
toroidal compactification, and since we already know the equivalent
type IIA theory in the latter case, one might imagine that the answer
to this question lies in taking an appropriate orbifold of the type
IIA string theory compactified on $K3$\cite{ASPIN,STROM}. While
this is precisely the procedure that was used in ref.\cite{FHSV}
to construct a dual
pair of theories with $N=2$ supersymmetry in four dimensions,
the situation is more subtle in our case.

For $D \le 5$ the construction is
simple\cite{ASPINTRIESTE}. For example, in five dimensions the dual theory
can be constructed by taking a type IIA theory compactified on
$K3\times S^1$, and then modding it by a $\ZZ_2$ action which includes a
shift on the $S^1$ by $\pi$, and whose action on $K3$ is such that
the $(0,0)$, $(2,2)$, $(0,2)$ and $(2,0)$ forms as well as
twelve out of the twenty
$(1,1)$ forms are invariant under this $\ZZ_2$. (The other eight
$(1,1)$ forms are mapped to their negatives.)
This reduces the rank of the resulting gauge group by eight.
This $\ZZ_2$ involution of $K_3$, which we call $\sigma$, will play an
important role in our subsequent discussion.
An explicit example of such an involution is given in the appendix.
Some of these models in four dimensions were constructed by
Ferrara and Kounnas\cite{FERRKOUN} in the fermionic description.

In six dimensions this construction breaks down. In fact,
at first sight it would seem to be
impossible to construct a type IIA string compactification in six
dimensions that is dual to the CHL compactification for the
following reason. There are three ways to get $N=2$ supersymmetry
in six dimensions from compactification of a type II string.
The first two possibilities are that all the supersymmetries
come from
the left (or right) sector of the world-sheet theory;
such theories are usually referred to as of type $(4_L,0_R)$ or
$(0_L,4_R)$, with the number $4$ representing the number of different
spin fields in the left or the right sector of the world sheet theory.
The remaining possibility is that half the supersymmetries
come from the left sector of the world sheet and the other half come
from the right sector. These theories are called $(2_L,2_R)$ type.
If the theory is of $(2_L,2_R)$ type, then using the same
conformal field theory as a background for the type IIB theory gives a
chiral theory in six dimensions. The massless field content
of a chiral $N=2$ theory in six dimensions is completely
determined by the requirement of anomaly cancellation\cite{SEIBERG}.
This implies that
we cannot a get a theory with a reduced rank
gauge group by using such conformal field theories for compactification
of the type IIA theory.
This argument does not rule out the possibility
of obtaining reduced rank theories from $(4_L,0_R)$ or $(0_L,4_R)$
compactification, since for such backgrounds both the type IIA and
type IIB theories give non-chiral theories.
However, if the theory is of $(4_L,0_R)$
type (say), then according to an argument of Banks and Dixon\cite{BANKSD}
the left sector of the world-sheet theory representing compact
dimensions is described by four free
superfields. This, in particular, implies that the only possible way to
get vertex operators representing
massless states from the Ramond-Ramond (RR) sector is to take the
product of the dimension $(1/4,0)$ spin field from the left sector of the
internal conformal field theory and
multiply it by a dimension $(0,1/4)$ spin field from the right sector of
the internal conformal field theory. (Recall that in six dimensions, a
massless state corresponds to a dimension (1/4,1/4) operator in the
RR sector of the internal conformal field theory.)
But since there is no supersymmetry coming from the right sector, there is
no dimension $(0,1/4)$ spin field in this theory. This shows that there
are no massless states arising from the RR sector of this
theory. Since under the duality transformation that maps heterotic string
theory to type IIA string theory, the gauge fields in the
heterotic string theory get mapped to the RR gauge fields in the type IIA
theory, we see that $(4_L,0_R)$ or $(0_L,4_R)$ compactifications
of the type IIA theory cannot give the required dual of the CHL
compactification in six dimensions.

Even though conventional compactification of
the type IIA string cannot give rise to the dual of the CHL
compactification in six dimensions, we can obtain the dual theory
by means of a type IIA compactification that involves
non-trivial RR background fields. To explain how this works, let us
return to the construction of the CHL compactification in six dimensions.
It consists of compactifying the heterotic string theory
on a four-dimensional torus and
modding out the resulting theory by a $\ZZ_2$ action that exchanges the
two $E_8$ lattices and simultaneously shifts one of the
circles of $T^4$ by half a period. Let us examine the action of this $\ZZ_2$
for the type IIA string theory compactified on $K3$
following a procedure similar
to the one adopted in ref.\cite{FHSV}. As already argued, the effect
of exchanging the two $E_8$'s can be represented by a $\ZZ_2$ involution
on $K3$, which we have denoted by $\sigma$. The result of a half period
shift along one of the circles can be represented in the type IIA theory
by a $\ZZ_2$ subgroup of the $U(1)$ gauge
group associated with the ten-dimensional gauge field $A_\mu$
originating in the RR sector\cite{WITTEN,FHSV}.\footnote{One way to
see this is to note that the U(1) gauge field in the heterotic
theory, associated with translation along the internal circle,
gets mapped to the RR gauge field $A_\mu$ of the type IIA theory
for a suitable choice of basis.}
Thus we conclude that if we
consider the $\ZZ_2$ orbifold of the type IIA theory compactified on $K3$
with $\ZZ_2$ acting on $K3$ as the involution
$\sigma$,  and also having a non-trivial embedding in the U(1) gauge group
involving the RR gauge field, the resulting theory is a good
candidate for the dual of the CHL compactification of the heterotic
string in six dimensions. From now on we shall refer to this new
type IIA string compactification as the twisted IIA theory.

The embedding of
$\ZZ_2$ in the U(1) gauge group implies that
all fields carrying even U(1) charge are even under the action of $\sigma$,
and all fields carrying odd U(1) charges are odd under the action of
$\sigma$. By a suitable gauge transformation that is singular on
$K3/\sigma$ we can make all the fields even under $\sigma$; but this
introduces background $U(1)$ gauge fields on $K3/\sigma$
whose effect is to give
a vev of $-1$ to any Wilson loop on $K3/\sigma$ that cannot be contracted
to a point without going through one of the orbifold points.
This implies that flux of the U(1) gauge field
is concentrated at the orbifold
points.  Embedding of the orbifold group
in the gauge group has been used many times before, notably in
usual orbifold compactification\cite{ORBIFOLD}, where one must
embed the spin connection in the gauge connection.
In the usual discussion of orbifolds the
gauge group originates in the NS sector, but in our case it originates
in the RR sector.

As is well known,
if we had not embedded $\ZZ_2$ in the U(1) gauge group,
and modded out the type IIA theory just by the action
of $\sigma$, the result would have been type IIA theory
compactified on another
$K3$. The point is that the modding out by $\sigma$ removes eight
massless vector multiplets from the untwisted sector corresponding to
the eight (1,1) forms that are odd under $\sigma$, but the twisted sector
gives back eight massless vector multiplets.\footnote{This is also
reflected in
the fact that if we blow up the orbifold points, we recover another
$K3$ surface which has twenty (1,1) forms.} The effect of introducing the
twist involving the RR sector gauge group is precisely to remove
these eight would-be massless multiplets from the twisted sector,
thereby reducing the rank of the gauge group by eight. The perturbative
spectrum in the untwisted sector is not affected by the embedding of $\ZZ_2$
into the gauge group, since all of these states are neutral under U(1).
Note that a Wilson line background for the RR gauge fields was also
introduced in ref.\cite{FHSV}, but there the complete perturbative spectrum
in the type IIA theory was insensitive to it. In the
present case, since the
involution $\sigma$ acts on $K3$ with fixed points, the
{\it perturbative spectrum of the type IIA theory in the twisted sector is
sensitive to the RR background gauge field.}

Unfortunately,
we cannot directly compute the spectrum of massless states
of the type IIA theory in the
presence of such background RR fields and demonstrate that the twisted
sector states disappear as claimed. We cannot even give an independent proof
that such a compactification of the type IIA string theory is consistent.
The problem, of course, is the lack of a suitable formalism for
describing RR backgrounds in terms of conformally invariant world-sheet
theories. In this sense the consistency of the CHL compactification
is the best proof of the consistency of such a type IIA string
compactification. However, as we shall now show, there is an
appropriate region in the moduli space of both these compactifications
where the effective low-energy theory is described by an
eleven-dimensional supergravity theory compactified on a five manifold
of large dimensions. The counting of massless
states in this effective theory  provides an `independent' way of
counting the number of massless states in both the theories.
This is done by exploiting another pair of duality conjectures
$-$ the duality between type IIA supergravity in ten dimensions and
eleven-dimensional supergravity on a circle\cite{TOWN,WITTEN}
and the duality between
heterotic string theory on $T^3$ and eleven-dimensional
supergravity compactified on $K3$\cite{WITTEN}.
Both of these dualities must be
regarded as dualities between effective theories, {\it i.e.} in the
appropriate limit the effective theory describing the string theory
is given by the corresponding supergravity theory. This implies that,
in the appropriate limits, both the heterotic string theory compactified
on $T^3\times S^1=T^4$ and type IIA theory compactified on $K3$
can be described by eleven-dimensional supergravity
on $K3\times S^1$, with both $K3$ and $S^1$ having large size.
In the supergravity theory, one of the vector fields in the
resulting six-dimensional theory comes from the component $G_{10,\mu}$
of the metric, where $10$ denotes the coordinate on $S^1$, and one
comes from dualizing the antisymmetric tensor field $C_{\mu\nu\rho}$
in six dimensions;
the other 22 vector fields, which
arise from the components $C_{mn\mu}$ of the antisymmetric tensor field
with $m,n$ denoting coordinates on $K3$, are in one-to-one
correspondence with the harmonic two-forms on $K3$.

Let us now consider a $\ZZ_2$ modding of the supergravity
theory that corresponds to the involution $\sigma$ on $K3$ and a shift on
$S^1$ by a half period. This $\ZZ_2$ action corresponds to the
same $\ZZ_2$ action that was used for constructing the CHL model from
the toroidally compactified heterotic string theory, and the $\ZZ_2$
that we used earlier in our construction of the twisted IIA string
compactification. Thus we claim that in some region of moduli
space both the
CHL compactification of the heterotic string and the twisted IIA string
compactification that we have proposed are
described at low energy by an effective theory consisting of
eleven-dimensional supergravity compactified on
$(K3\times S^1)/\ZZ_2$, where the $\ZZ_2$ acts as the involution $\sigma$
on $K3$ and a half-period shift on $S^1$.
Since this $\ZZ_2$ action on $K3\times S^1$ has no fixed points,
the supergravity theory is well-defined on the
quotient manifold as an effective theory, and we can reliably count the
total number of massless states in the theory by simply counting the
massless states in this effective supergravity theory. Since there are
no twisted states in the supergravity theory, the result of the
$\ZZ_2$ action is simply to eliminate the vector multiplets
corresponding to (1,1) forms that are odd under $\sigma$.
As already mentioned,
there are eight such (1,1) forms, and hence the resulting theory has the
rank of its gauge group reduced by eight.

We can learn more about the twisted IIA theory
by considering compactification
to five dimensions. Before the $\ZZ_2$ modding, we can consider
four equivalent descriptions of the theory: 1) the eleven-dimensional
theory compactified on $K3\times S^1_a \times S^1_b$, where we have
added the labels $a$ and $b$ to distinguish the two circles; 2) type
IIA theory on $K3 \times S^1_a$ (we shall denote this theory
by IIAb); 3) type IIA theory on $K3\times S^1_b$ (this theory
will be denoted by IIAa); and 4) heterotic string theory on
$T^3\times S^1_a \times S^1_b$. There are also equivalent type IIB
string compactifications, which we shall not discuss. Let us denote by
the superscripts $\SG$, $\iia$, $\iib$ and $\het$ the fields in the
supergravity, IIAa, IIAb, and heterotic string theory, respectively.
We also denote by $V_M$ the volume of a manifold $M$,
by $R_i$ the radius of the circle $S_i^1$, and by $\Phi_n$ the $n$-dimensional
dilaton. Then the relevant moduli fields setting the
scales in these four theories are as follows:

\begin{itemize}

\item{Supergravity theory: $V^\SG_{K3}$, $R^\SG_a$, $R^\SG_b$.}

\item{IIAa theory: $\Phi^\iia_{10}$, $V^\iia_{K3}$, $R^\iia_b$.}

\item{IIAb theory: $\Phi^\iib_{10}$, $V^\iib_{K3}$, $R^\iib_a$.}

\item{Heterotic theory: $\Phi^\het_7$, $R^\het_a$, $R^\het_b$.}

\end{itemize}

Following ref.\cite{WITTEN} we can find the relationship between
the moduli of different theories. They are given by
\ben \label{ee1}
V^\SG_{K3} &=& \exp({4\over 3}\Phi^\het_7) \nonumber \\
&=& V^\iia_{K3} \exp(-{4\over 3} \Phi^\iia_{10}) \nonumber \\
&=& V^\iib_{K3} \exp(-{4\over 3} \Phi^\iib_{10}) \, ,
\een
\ben \label{ee2}
R^\SG_a &=& \exp(-{2\over 3}\Phi^\het_7) R^\het_a \nonumber \\
&=& \exp({2\over 3} \Phi^\iia_{10}) \nonumber \\
&=& R^\iib_a \exp(-{1\over 3} \Phi^\iib_{10}) \, ,
\een
\ben \label{ee3}
R^\SG_b &=& \exp(-{2\over 3}\Phi^\het_7) R^\het_b \nonumber \\
&=& R^\iia_b \exp(-{1\over 3} \Phi^\iia_{10}) \nonumber \\
&=& \exp({2\over 3} \Phi^\iib_{10}) \, .
\een
If $G_{\mu\nu}$ denotes the five-dimensional metric ($0\le \mu,\nu
\le 4$) then the metrics of the different theories are related as
follows:
\be \label{ee4}
G^\SG_{\mu\nu} = e^{-{4\over 3} \Phi^\het_7} G^\het_{\mu\nu}
= e^{-{2\over 3} \Phi^\iia_{10}} G^\iia_{\mu\nu}
= e^{-{2\over 3} \Phi^\iib_{10}} G^\iib_{\mu\nu}\, .
\ee
Let $C_{MNP}$ denote the three form field of the eleven-dimensional
supergravity theory, $A_M$ denote the U(1) gauge field arising in
the RR sector of the type IIA string theory, and $B_{MN}$ be the
usual antisymmetric tensor field of the IIA theory.
Then the relationship between
different gauge fields in five dimensions is given by,
\ben \label{ee5}
G^\SG_{a \mu} \sim G^\het_{a\mu} \sim G^\iib_{a\mu} \sim
A^\iia_\mu \nonumber \\
G^\SG_{b \mu} \sim G^\het_{b\mu} \sim
A^\iib_\mu \sim G^\iia_{b\mu} \nonumber \\
C^\SG_{ab\mu} \sim \wt B^\het_{\mu} \sim B^\iib_{a\mu} \sim
B^\iia_{b\mu}\, ,
\een
where $\wt B_\mu$ denotes the dual of the antisymmetric tensor
field in five dimensions. The symbol $\sim$ in the above relations
signifies that we have ignored normalizations. The subscripts $a$ and $b$
refer to the circles $S_a^1$ and $S_b^1$.

A fundamental type IIAa string wrapping around the circle
$S_b^1$ corresponds to a state with a non-zero $B^\iia_{b\mu}$
charge. Let us normalize $B^\iia_{b\mu}$ so that this state carries
unit charge. Similarly, we shall normalize $B^\iib_{a\mu}$ in such a
way that a fundamental type IIAb string wrapped around the circle
$S_a^1$ carries unit $B^\iib_{a\mu}$ charge. From eq.\refb{ee5} we
see that the field $B^\iib_{a\mu}$ in the IIAb theory gets mapped
to the field $B^\iia_{b\mu}$ in the type IIAa theory. Due to the
symmetry of the exchange $a\leftrightarrow b$, the
relative normalization between these two fields must be $\pm 1$.
Thus we learn that the fundamental type IIAb string wrapped
around the circle $S_a$ represents the same state as the
fundamental type IIAa string wrapped around the circle $S_b$.
This can be further checked by comparing the masses of these states;
the fundamental IIAa (IIAb) string, wrapped around $S_b^1$ ($S_a^1$)
has mass $R^\iia_b$ ($R^\iib_a$) measured in the metric
$G^\iia_{\mu\nu}$ ($G^\iib_{\mu\nu}$). Using
eqs.\refb{ee1}--\refb{ee4} one can easily check that these correspond
to identical masses.\footnote{Another way to understand this
phenomenon is to note that both these states can be regarded as
the 11-dimensional supermembrane wrapped around $S_a^1\times S_b^1$.}

So far we have only collected the relevant results of
ref.\cite{WITTEN}. Let us now consider modding out each of the four
theories by $\ZZ_2$, where $\ZZ_2$ acts as the involution $\sigma$
on $K3$ and a shift by $\pi$ on $S_a^1$. On the heterotic side
this corresponds to CHL compactification.
For the IIAb theory, this
corresponds to the compactification of an usual type IIA theory
on $(K3\times S^1_a)/\ZZ_2$.
For the IIAa theory, on the other hand, it
corresponds to the twisted IIA theory compactified
on the circle $S^1_b$.
Thus the twisted
IIAa theory compactified on the circle $S^1_b$ must be
(non-perturbatively)
equivalent  to the IIAb theory compactified on the manifold
$(K3\times S^1_a)/\ZZ_2$. In particular, the spectrum of these two
theories must agree. Now, the IIAb theory compactified on $(K3\times
S^1_a)/\ZZ_2$ has states with half-integer
$B^\iib_{a\mu}$ charge\cite{WENWIT}, since there can be winding
states of the string that wind half way around $S^1_a$, and
at the same time start and
end on points of $K3$ related by the involution $\sigma$. Thus
the twisted IIAa theory compactified on the circle $S^1_b$ must
also contain states with half integer $B^\iia_{b\mu}$ charge. As has
been stated before, in this theory states with integer
$B^\iia_{b\mu}$ charge  correspond to fundamental strings wound
around the circle $S^1_b$. But now we have a puzzle: since
the string cannot have half a winding around $S^1_b$, there is no way
to get states with half a unit of $B^\iia_{b\mu}$ charge by winding
fundamental strings around $S^1_b$. The only resolution to this puzzle
is that the original twisted IIAa theory in six dimensions
has a solitonic string that {\it
carries half the density of $B_{\mu\nu}$
charge of a fundamental string.}\footnote{From the eleven-dimensional
viewpoint such a string would correspond to a solitonic membrane labelled
by surface coordinates $(\sigma_1, \sigma_2)$, with $\sigma_1$ pointing
towards a particular direction in the 5-dimensional non-compact
space, and $\sigma_2$ lying along a closed curve $C$ in
$(K3\times S^1_a)/\ZZZ_2$, with two ends of of $C$ being related by the
$\ZZZ_2$ transformation in $K3\times S^1_a$. Classically the minimum
energy configuration for such a membrane would correspond to the choice
of $C$ which lies along $S^1_a$ spanning a length $\pi$, and sits
at one of the eight fixed points of $\sigma$ on $K3$. Viewed as a
string in six dimensions, this will have half the tension of a
fundamental string which is obtained by wrapping the membrane fully
around $S^1_a$.}
Winding of this string along $S^1_b$
would then produce a state with half a unit of $B^\iia_{b\mu}$ charge.

To summarize, we have shown that the dual of the six-dimensional CHL
compactification of the heterotic string theory is given by the
compactification of the type IIA theory on a $K3/\ZZ_2$ orbifold,
where the $\ZZ_2$ action involves an involution of $K3$, and also
has an embedding in the U(1) gauge group arising from the RR sector
of the ten-dimensional string theory.
Geometrically
this represents flux associated with this U(1) gauge field concentrated
at the orbifold points. This construction shows that it is possible to
construct non-trivial background for the propagation of type IIA
string with non-vanishing background fields associated with the
RR sector. This new six-dimensional string theory contains solitonic
strings which carry half the density of $B_{\mu\nu}$ charge as the
fundamental string.

{\underbar{Acknowledgement}}
We wish to thank A. Dabholkar, J. Harvey,
K.S. Narain, and A. Strominger for valuable discussions.
The work of J.H.S. is supported in part by the U.S. Department of
Energy Grant No. DE-FG03-92-ER40701. A.S. would like to thank the
theory group at Caltech for hospitality during the course of this
work.

\appendix

\renewcommand{\theequation}{\thesection.\arabic{equation}}

\sectiono{Example of $Z_2$ Involution on $K3$}

In this appendix we shall construct an example of
the involution $\sigma$ that eliminates eight of
the $(1,1)$ forms.
For this purpose let us consider the $K3$ surface described by the quartic
polynomial equation in $\CC {\rm P}^3$:
\be \label{one}
\sum_{i=1}^4 (z^i)^4 =0 ,
\ee
where $z^i$ are the homogeneous coordinates in $\CC {\rm P}^3$. Let us denote
by $\sigma$ the involution:
\be \label{etwo}
z^1\to -z^1, \quad z^2 \to -z^2, \quad z^3\to z^3, \quad z^4\to z^4\, .
\ee
It can be easily verified that it preserves the $(2,0)$ and $(0,2)$
forms on $K3$. There are eight fixed points on $K3$ under the action
of $\sigma$, described by the points
\ben \label{ethree}
z^1=z^2=0, \qquad {z^3\over z^4}= e^{(2k+1)i\pi/4}, \quad k=0,1,2,3,
\nonumber \\
z^3=z^4=0, \qquad {z^1\over z^2}= e^{(2k+1)i\pi/4}, \quad k=0,1,2,3.
\een
If $n_+$ and $n_-$ denote the number of $(1,1)$ forms on $K3$ that are
even and odd under $\sigma$, respectively, then by Lefschetz fixed-point
theorem\cite{LEFS}, we have
\be \label{efour}
8 = 4 + n_+ - n_-\, ,
\ee
where the number 4 on the right-hand side represents the contribution from
the (0,0), (2,2), (2,0) and (0,2) forms, all of which are even under
$\sigma$.
Since,
\be \label{efive}
n_+ + n_- = 20\, ,
\ee
we get
\be \label{esix}
n_+=12, \qquad n_-=8\, .
\ee
This shows that eight out of the twenty harmonic (1,1) forms are odd under
$\sigma$, as required.

\end{document}